\documentclass[conference]{IEEEtran}
\IEEEoverridecommandlockouts
\usepackage{cite}
\usepackage{amsmath,amssymb,amsfonts}
\usepackage{algorithmic}
\usepackage[linesnumbered,ruled]{algorithm2e}
\usepackage{graphicx}
\usepackage{textcomp}
\usepackage{xcolor}
\usepackage{array}
\usepackage{wrapfig}
\usepackage{multirow}
\usepackage{tabu}
\usepackage{url}
\def\BibTeX{{\rm B\kern-.05em{\sc i\kern-.025em b}\kern-.08em
    T\kern-.1667em\lower.7ex\hbox{E}\kern-.125emX}}
    
\newif\ifdraft
\draftfalse
\ifdraft
 \newcommand{\eunsungnote}[1]{ {\textcolor{red} { ***Eunsung: #1 }}}
 \newcommand{\shahzadnote}[1]{ {\textcolor{blue} { ***Shahzad: #1 }}}
\else
 \newcommand{\eunsungnote}[1]{}
 \newcommand{\shahzadnote}[1]{}
\fi
\newcommand{\ignore}[1]{}

\begin{document}

\title{FLIP-FLexible IoT Path Programming Framework for Large-scale IoT\\
{\footnotesize \textsuperscript{}}
\thanks{\ignore{Identify applicable funding agency here. If none, delete this.}}
}

\author{\IEEEauthorblockN{Shahzad}
\IEEEauthorblockA{\textit{Department of Electronics and Computer Engineering} \\
\textit{Hongik University, Korea}\\
shahzad@mail.hongik.ac.kr}
\and
\IEEEauthorblockN{Eun-Sung Jung\eunsungnote{typo in my name}}\shahzadnote{edited}
\IEEEauthorblockA{\textit{Department of Software and Communication Engineering} \\
\textit{Hongik University, Korea}\\
ejung@hongik.ac.kr}
}

\maketitle

\begin{abstract}
With the rapid increase in smart objects forming IoT fabric, it is inevitable to see billions of devices connected together, forming large-scale IoT networks. This expeditious increase in IoT devices is giving rise to increased user requirements and network complexity. \eunsungnote{The following sentence is grammatically wrong.} Collecting big data from these IoT devices with optimal network utilization and simplicity is becoming more and more challenging.\shahzadnote{$in\rightarrow is$} This paper proposes FLIP- FLexible IoT Path Programming Framework for Large-scale IoT. The distinctive feature of FLIP is that it focuses on the IoT fabric from the perspective of user requirements and uses SDN techniques along with DPI technology to efficiently fulfill the user requirements and establish datapath in the network in an automated and distributed manner. FLIP utilizes SDN structure to optimize network utilization through in-network computing and automated datapath establishment, also hiding network complexity along the way. We evaluated our framework through experiments, and results indicate that FLIP has the potential to fulfill user requirements in an automated fashion and optimize network utilization.
\end{abstract}

\begin{IEEEkeywords}
 Software-defined Networks (SDN), Deep Packet Inspection(DPI)\eunsungnote{Remark: I believe that our work can be done not only with DPI but also with P4.}
\end{IEEEkeywords}

\section{Introduction}
\eunsungnote{I'll get back to the introduction section after all the other sectionis are done.}
The Internet of Things (IoT) is an increasingly important and revolutionary paradigm for enabling the world of smart objects (SOs). An estimated 50 billion devices will be connected to the Internet by 2020 \cite{cisco:whitepaper}.  With this expansion, the user requirements for an IoT system are also increasing e.g. datapath selection, computation, delay, rate, jitter, and energy efficiency etc. 

Software defined networking (SDN) brings about innovation, simplicity in network management, and configuration in network computing \cite{SDN:wirelesssurvey}. In recent times researchers have provided SDN based solutions to different IoT problems \cite{muppet}, \cite{fog_routing_2018}. But SDN also has its limitations. It is largely restricted to L2-L4 protocols \cite{Extendingopenflow} \cite{atlas} \cite{Mekky} \cite{Bhat} \cite{DT} . It means that we cannot process the packets’ data above L4. There have been techniques proposed to tackle this limitation. Udechukwu \textit{et al.} \cite{Extendingopenflow} uses a middlebox as a DPI engine to apply traffic engineering to get better video stream. Application based QoE support is proposed in \cite{Bhat} ,it also focuses on video streaming quality. Atlas framework \cite{atlas} employs a machine learning (ML) based traffic classification technique. Bui \textit{et al.} \cite{DT} came up with a generic interface to extend the Open vSwitch to recognize custom protocols but this interface is very limited and does not allow users to manipulate packet payload.  Extended-SDN architecture \cite{Mekky} uses an extended table in the Open vSwitch called “application table” to keep track of packets based on L7 header information. It is worth mentioning that non of the above solutions are proposed for IoT systems. In contrast to the above techniques, our framework is designed in consideration to large-scale IoT. FLIP not only identifies L4-L7 headers but also processes the application layer payload which in most IoT cases can be helpful to reduce network resource utilization.

In this paper, \eunsungnote{use paper instead of article throughout this paper.}\shahzadnote{$article\rightarrow paper$} we present a flexible IoT path programming framework that applies SDN techniques with External Processing Boxes (EPBs). The embedded DPI engine inside an EPB analyzes, processes, and forwards packets such that big IoT data is efficiently routed to the destination. 
\ignore{We have focused on key user requirements to be used for efficient datapath selection and routing. Since most of these requirements depend on the application layer data, we have introduced a DPI engine in our framework to analyze and process L4 to L7 layer packet payload. }
Our contributions in this paper are three-fold. First, we propose an SDN-based IoT datapath programming framework using EPBs. Second, we define user requirements formally and design python-base script programming syntax to incorporate user requirements. Third, we propose an efficient data path computation algorithm that outputs network resource saving datapath while satisfying user requirements. Finally, the efficacy of the proposed algorithm is shown though microbenchmarking measuring packet counts at each switch.

\eunsungnote{The following paragraph should be changed according to the section modifications}\shahzadnote{changed}
The rest of this paper is organized as follows. Section \ref{Problem-Statement} details the problem statement and the pertaining target network and user requirements. Section \ref{Architecture} explains the overall architecture of FLIP. Section \ref{syntax} provides an overview of the User command syntax of FLIP. Section \ref{datapath installation} details FLIP's flexible datapath installation and heuristics used for the datapath selection. Experiments to evaluate the framework performance are detailed in Section \ref{Experiments}. Section \ref{Related-work} provides a review of the related work. Section \ref{conclusion} concludes the paper. 
\eunsungnote{The conclusion section is indispensable in the paper. Add a conclusion section.}\shahzadnote{Conclusion Section added}

\begin{figure*}[t]
\centerline{\includegraphics[scale=0.45]{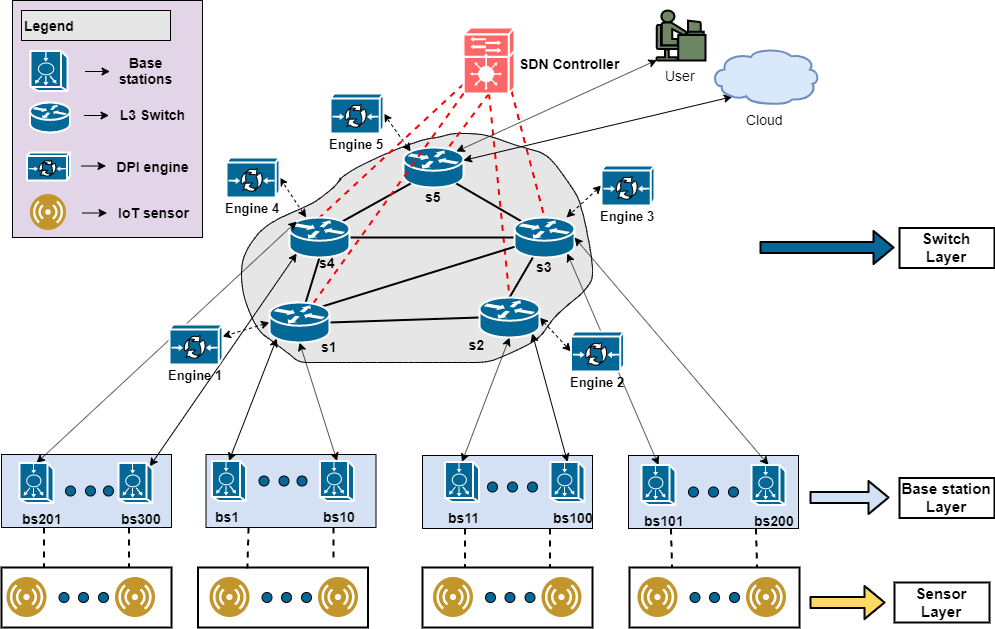}}
\caption{Target Network}
\label{fig-target-network}
\end{figure*}
\section{Problem Statement\eunsungnote{change label for this. User requirements $->$problem statement}}\shahzadnote{done} \label{Problem-Statement}

In general, IoT data processing takes place at the servers or clouds designated by users. These days, most cloud computing providers provide IoT data processing services in their clouds. In the case of Amazon AWS,  IoT devices send their data to the IoT Hub, and cloud computing resources subsequently process the data.

However, when there are tremendous numbers of IoT devices and users, network bottleneck and storage/computing resource wastage due to unnecessary data processing may not satisfy many users' requirements properly. 
Suppose that we need the average value of data from millions of IoT devices. If all IoT devices send their data to servers or clouds, network traffic bottleneck or lack of computing resources for handling massive IoT data may hinder proper IoT data processing.
Instead, if we may compute the average value in the network, and send the computed results, we can save network/computing resources, and also process IoT data properly. 
We propose efficient in-network processing of IoT data, which will reduce the amount of data arriving at servers or clouds and subsequently make it possible to cope with massive IoT data processing while meeting each user's requirements. 
We define our target networks and problem in a formal way in the following subsections.
\subsection{Target Networks}
\ignore{Choosing a target network in IoT is not easy because of the heterogeneous nature of IoT networks and varying topology structure.}
IoT networks and associated topology structures and access technology for communication vary depending on applications (e.g., smart home, smart grid, VANET, etc.). 
Hanes \textit{et al}.\cite{hanes_iot_2017}  summarizes the commonly used topologies in IoT as star, mesh, partial mesh, and peer to peer. Huang \textit{et al}.\cite{huang_topology_2017}\eunsungnote{Be consistent in using name or authors. Previously you used the last name of the first author}\shahzadnote{got it.} proposes a three-layered (base station layer, relay node layer, and sensor layer) energy-efficient topology
structure for IoT. A recent study also presents a three-tiered
(Cloud tier, Fog Tier, and IoT tier) topology structure for the
IoT \cite{fog_routing_2018}. \eunsungnote{The following sentence is too vague. Rewriting is required. You chose what models from mentioned work?}\shahzadnote{Changed the sentence, added a sensor layer in Figure 1 as well} Based on these recent studies we chose a three layered (Switch layer, Base station layer and Sensor layer) topology structure similar to \cite{huang_topology_2017}. Switch layer follows a partial-mesh topology model \cite{hanes_iot_2017}. Base station layer and sensor layer follow the same hierarchical model as mentioned in \cite{huang_topology_2017} and \cite{fog_routing_2018}. we have selected a target network that can adapt to the referred topology structures easily.

Fig. \ref{fig-target-network} shows our target network model. The SDN controller connects to L3-switches\eunsungnote{ovs is just for our simulation environment. In general, this should be hardware switches.refer to the below comments on  Fig. 1.}\shahzadnote{changed ovs to L3-switch. also changed in figure 1.} and has an overall visibility of the network. The L3-switch can be a software switch or a hardware switch. In case of software switch like Open vSwitch (OVS), both OVS and DPI engine can run on the same machine. While in hardware switch's case, we assume a programmable switch (e.g. Cisco's Nexus 3000) with internal or external DPI engine. Each switch connects to several base stations that are responsible for collecting data from the sensor layer. In the literature, the term, base station, is interchangeably used with access point, sink, and gateway point \cite{hanes_iot_2017}, \cite{huang_topology_2017}, \cite{fog_routing_2018}.
\ignore{In the IoT community, there are different terminologies used for this base station e.g. access point, sink, gateway point etc \cite{hanes_iot_2017}, \cite{huang_topology_2017}, \cite{fog_routing_2018}.} 
We suppose that each switch has a connected DPI\eunsungnote{Where did you define DPI? If you didn't, explain here.}\shahzadnote{full form for the acronym is given before, but definition is in Architecture, added a reference for it here} engine that can process packets passing through the switch. DPI engine is explained in Section \ref{DPI engine}. We will refer to this target network in the following sections to explain the examples. 

\eunsungnote{You may elaborate switch. switch can be a software switch or a hardware switch. For the sw switch, ovs and dpi are running in the same machine. For the hw switch, assume a programmable switch with dpi inside or outside.}\shahzadnote{done}

\subsection{User requirements}
\ignore{As mentioned in section I, our objective is to consider various user requirements that are common in an IoT environment for datapath establishment and routing.}
Our goal is to establish appropriate datapaths for massive IoT data with efficient in-network data processing and routing while meeting various user requirements. FLIP, our proposed framework, utilizes SDN and DPI techniques for datapath establishment and in-network processing, respectively. For clear understanding of user requirements, we articulate following user requirements one by one in the next sections. \ignore{In this article we want to show that SDN techniques can be utilized to achieve this objective. FLIP is designed in consideration (but not limited) to the following user requirements in IoT environments}

\begin{itemize}
    \item Data type (e.g., scalar or vector)
    \item Coverage (e.g., area)
    \item Delay (e.g., \textless 10ms)
    \item Rate (e.g., 1s)
    \item Jitter (e.g., \textless 5ms)
    \item Computations (e.g., min(), max())
\end{itemize}

\subsubsection{Data type}
\ignore{Data type refers to the type of data being provided by the sensors to base stations.} 
Data type refers to what data users want to collect from sensors. 
\eunsungnote{data type is given in command syntax? Isn't this internal stuff? Then we should not include under user requirements. needs to be rewritten or relocated somewhere else.}\shahzadnote{It was not part of syntax but I added some detail providing syntax base for including this into user requirements, have a look}
\textbf{scalar} indicates that only one sensor is providing the data (e.g., temperature reading from a single sensor). 
\textbf{vector} indicates that multiple sensors are sending data at a given instance. 
\textbf{matrix} indicates that multiple vectors are selected to send data for a certain period of time. 
In our paper we have assumed that the base stations are resourceful nodes and are using one of the common IoT protocols i.e. MQTT, CoAP, HTTP, WebSockets. 
This means base stations are capable of encapsulating different types of data into a single payload. For example MQTT is a famous IoT protocol \cite{hanes_iot_2017}. It is a publish-subscribe based protocol commonly used in IoT environments. There is a field called $packet\ type$ and $packet\ length$ in it's packet header. These field values can be utilized to identify the packet payload. CoAP is another common IoT protocol \cite{hanes_iot_2017}. It has a different packet structure compared to MQTT. To deal with such protocols, our DPI engine is capable of dissecting the protocol headers and inspect the packet payload that is being provided by the base station. The user has two types of command syntax at his disposal, automated and manual (explained later). In the automated case, the protocol is identified by the framework itself. While in the manual approach, user provides information about the IoT protocol being used. The framework then uses this information to match incoming packet on the selected protocol only. In this way we can manipulate different data types. More detail on DPI engine and command syntax is in the Section \ref{DPI engine} and \ref{syntax} respectively.

\subsubsection{Coverage}
In many IoT scenarios such as communication between hospitals in one or more regions\cite{IoTstandards}, or gathering sensory weather data from multiple sites, coverage is an important user requirement.
 \ignore{Although it has not gotten much attention when it comes to user's simplicity in selecting the coverage area for collecting information.} 

We introduce a naming translation service in FLIP that will help the user to identify IoT device addresses \eunsungnote{user doesn't care about base stations, but IoT device addresses.}\shahzadnote{ok. changed it} in relation to their coverage, which has been identifiable by network addresses so far. For example if the user wants to get IoT data from a certain region in Korea like Seoul, instead of requesting it by supplying network addresses for IoT devices residing in that area, the user is only required to enter ``Seoul" and FLIP will translate this information into network readable addresses. In another scenario the user may want to get the inventory information from all the hospitals in Chicago. \eunsungnote{better not to use Sejong, which is not well-known international city.}\shahzadnote{Right, Sejong$\rightarrow$ Chicago}
The user will enter ``Chicago" instead of providing a list of hospitals to the system to get the required information. This is done by FLIP's translation module explained in Section \ref{Architecture}.
\eunsungnote{regarding coverage, add some sentences comparing with Amazon IoT, https://aws.amazon.com/iot-core/features/, especially message broker}\shahzadnote{added} This translation feature of FLIP is similar to Amazon AWS IoT message broker which acts as a mediator between clients and the AWS IoT core. Clients (users and IoT devices) communicate with each other through this message broker using protocols like MQTT, WebSockets and HTTP. Clients(IoT devices) send data by publishing messages on a topic and clients(users) receive messages by subscribing to a topic. The message broker is responsible for maintaining mapping between publishers and subscribers based on the message topic. FLIP in a similar fashion maintains mapping between IoT device addresses and their physical location. In this paper we assume that user wants to collect the IoT data from all the IoT devices connected to a base station. Hence we refer to base stations as source nodes where needed.

\subsubsection{Delay}
\begin{figure}[t]
\centerline{\includegraphics[scale=0.32]{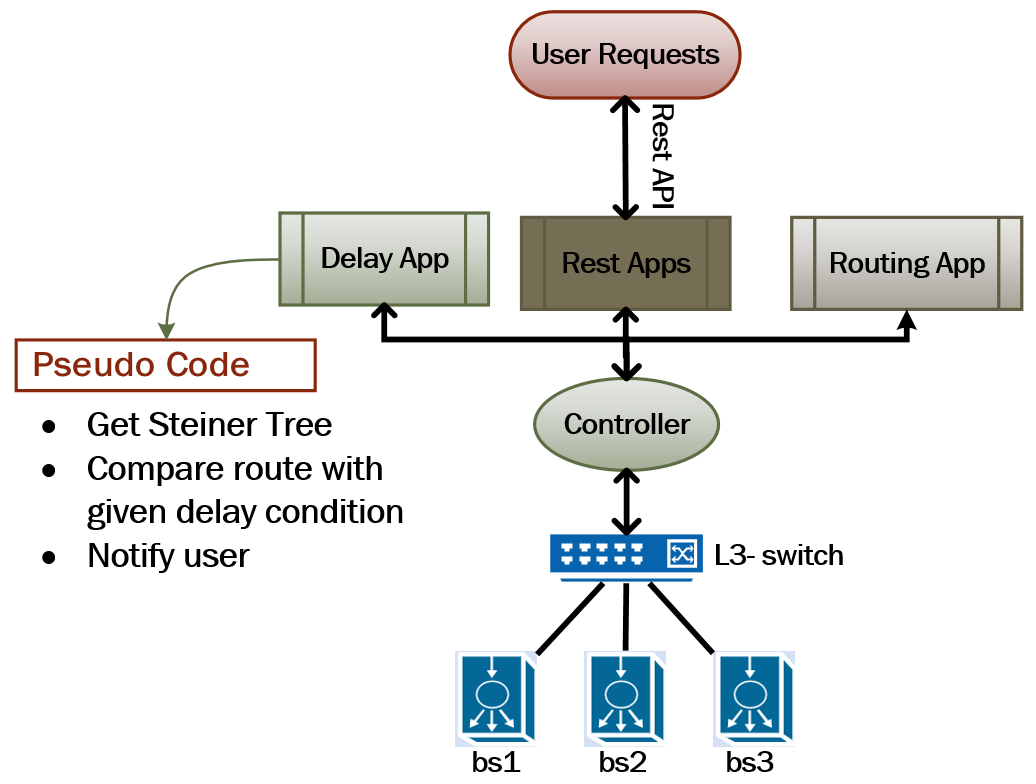}}
\caption{Implementation of delay requirement}
\label{fig-delay}
\end{figure}

The delay requirement limits the maximum time taken from data sampling at base stations to user designated destination. 
When a user gives a predefined delay value, FLIP is supposed to find a route for the user that meets the delay requirement. 
For this we have a RYU application as shown in Fig. \ref{fig-delay} dedicated to check this requirement. \eunsungnote{Remark: we need more sophisticated algorithm for this later.} This application looks at the Steiner tree created for the datapath (see Section \ref{datapath installation}). It compares the delay for path to be taken with user delay requirement and determines if it is possible or not and then notifies the user respectively. The translation module checks the user request and if delay parameter is provided, it invokes the delay application through a REST call.
\eunsungnote{typo in Fig. 2. Seiner Tree. api$->$API.}\shahzadnote{corrected}
\subsubsection{Rate}
\begin{figure}[t]
\centerline{\includegraphics[scale=0.44, trim= 0 5 0 0,clip]{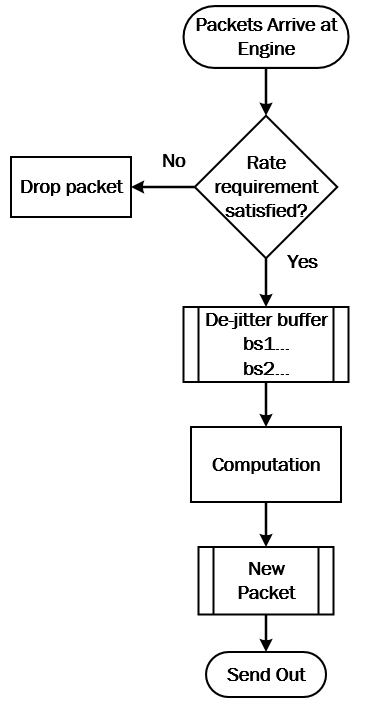}}
\caption{Implementation of rate, jitter, and compute requirements}
\label{fig-rate}
\end{figure}

In most IoT scenarios, the sensor data is published either at a fixed rate\eunsungnote{What do you mean? Physically, there should be sampling rate.}\shahzadnote{that is true. in low level sensors it is fixed but in advance smart objects (SOs), we have programmatic control over this to some extent, hence the term "predefined".} or at a predefined interval. The rate requirement allows users to select the IoT data collection rate. For example, the base stations are publishing data at the rate of 100ms but the user wants the data to be only received at the  rate of 1sec. To achieve this we have added the rate requirement check in FLIP. 
Fig. \ref{fig-rate} shows how the rate requirement can be implemented. As packets arrive at the DPI engine, the rate requirement is checked, and packets are dropped if they do not meet the requirement,  otherwise, they are forwarded to the next hop. This is done by the Translation module\eunsungnote{This is ambiguous. The actual dropping happens at the switch.}\shahzadnote{switches drop packets but here we want to drop them in comparison to the user rate value which is received at the engine. Also engine can drop the packets too. We redirect the packets from base station on the 1st hop to the engine. check the rate at engine and decide} that configures the DPI engine with the help of engine configuration file (Section \ref{Architecture}) 

\eunsungnote{Is there any colorcode for the flowchart? If not, use white background color for boxes.}\shahzadnote{white background added}

\subsubsection{Jitter}
\eunsungnote{The definition of jitter in our context is different from the usual jitter. Usual: variation of latency at the destination. Ours: difference between sensor data timestamps in addition to usual definition. For example, rates are set to 1s, there is a possibility that 1s timestamp difference among packets from IoT devices. How do you handle this in you de-jitter buffer? If you have thoughts on this, elaborate.}\shahzadnote{..} Packet jitter is one of the most important QoS metrics of real time services \cite{IoT-traffic-Qos}. In general, Jitter is the variation of delay/latency in arriving packets at the destination. In our framework we consider jitter as difference between sensor data timestamps arriving at a DPI engine for processing. For example the user requires the temperature data of multiple sensors from a base station and this data arrives with jitter at an engine for processing. The DPI engine should check the arrival offset between packets and consider only those packets that meet a certain predefined jitter threshold. For this we maintain a de-jitter buffer (see Fig. \ref{fig-rate}) at the DPI engine for each incoming port. This buffer value is also controlled through the Translation module and engine configuration file (Section \ref{Architecture}).  The ITU-R has recommended 25 ms jitter as an acceptable value for the delay variation \cite{VoIP-handbook}. So FLIP expects a value of $0\sim 25ms$ from the user if jitter is selected as a requirement.

\subsubsection{Computation}
Computation is the crucial user requirement for our purpose of in-network data processing to reduce IoT data packets dramatically and relieve burdens on the server/cloud side. Computation is the main reason to add DPI engine into the framework. In most IoT scenarios the user is only interested in processed data. \eunsungnote{I don't understand how PHEME is related to our work from the above sentence.}\shahzadnote{citing PHEME here is to support the predecessor sentence (interest in processed data). other than that PHEME is mentioned in the related work} In our framework, the DPI engine can read and process L4-L7 data from the incoming packets. For now, FLIP supports basic computation operations such as $min()$, $max()$, $sum()$, $sub()$, $avg()$, and $mul()$. Syntax detail on how a user can associate a compute operation with incoming data is explained in Section \ref{Architecture}.

\section{Overall Architecture} \label{Architecture}
\ignore{There are three major components in our framework as shown in Fig. \ref{fig-architecture} and explained in the following subsections.}
We describe the overall architecture of our framework to serve user requests for IoT data collection and processing. Our framework is composed of three major modules, Translation module, SDN module, and EPB.

\subsection{Translation module}
\eunsungnote{The following sentence is not meaningful at all. comment it out.} \ignore{The translation module is a very important component of FLIP}. \eunsungnote{The following sentence is redundant with the next sentence. comment it out.} \ignore{It connects the user with the framework.}\shahzadnote{done} 
The translation module translates user requests and passes the translated output to FLIP. The translation module is provided to users as a python library. The user can simply import this module into a python script and write requests using the provided syntax (see Section \ref{syntax}). This module communicates with SDN module and EPB via REST APIs. More specifically, it uses python ``requests" library. In summary, when a user sends a request in python, the request is translated by this module into multiple REST requests which will then be shared with SDN module and EPB in a distributed manner. This module is also responsible for naming translation for the coverage requirement and datapath selection which is described in detail in Section \ref{datapath installation}.

\subsection{SDN module}
The SDN module consists of:
\begin{itemize}
    \item RYU SDN controller
    \item OpenFlow compatible (hardware/software) switches 
    \eunsungnote{ovs=open virtual switch, not openflow virtual switch. In addition, this part should be generic. replace by "OpenFlow compatible software/hardware switches."}\shahzadnote{done}
    \item RYU applications \ignore{(discussed later)}
    \begin{itemize}
        \item delay application
        \item ofctl\_rest application
        \item rest\_topology application
        \item routing\_ryu application
    \end{itemize}
\end{itemize}   
Each switch has one or more connected base stations and one EPB \eunsungnote{use full name}\shahzadnote{changed back to acronym, because I have given the full form at 1\textsuperscript{st} occurrence}. These base stations are assumed to be resourceful nodes that can communicate with common IoT protocols (e.g., MQTT and CoAP). \ignore{Braun et al in \cite{braun} has detailed different options for the southbound protocol in SDN among which OpenFlow is the most common.} Even though there are many other options for the southbound protocol for communication between a controller and switches, we use OpenFlow, the most prevalent one recently. Regarding the northbound communication between a controller and applications, we use REST APIs. 

\subsection{External Processing Box (EPB)\eunsungnote{use full name.}}
\shahzadnote{Ok}\label{DPI engine}
The EPB consists of:
\begin{itemize}
    \item Flask web application \eunsungnote{If I remember correctly, flask is an off-the-shelf software. If that is the case, cite it and explain a little bit more.}
    \item Engine configuration file \eunsungnote{Be consistent with Capitalization. Configuration-\textgreater configuration. same for the below line.} \shahzadnote{changed the above. following symmetry for the below because of the acronym }
    \item Deep Packet Inspection (DPI) engine 
\end{itemize}

\begin{figure}[htbp]
\centerline{\includegraphics[scale=0.8, trim= 0 0 0 0,clip]{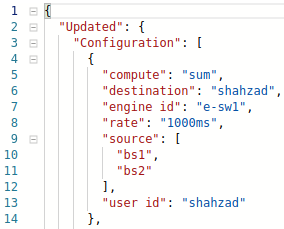}}
\caption{Engine configuration file snippet}
\label{fig-engineconfig}
\end{figure}

Flask is a lightweight wsgi (web server gateway interface) web application framework \cite{flaskwebsite}. This framework is used to make web server applications that can be accessed through REST APIs. We developed a flask application that provides a REST interface to the Translation module  to manage the engine configuration file. 
The engine configuration file maintains the DPI engine settings for each request. A snippet of engine configuration file is shown in Fig. \ref{fig-engineconfig}. As shown in the figure, this configuration file is showing the current configuration of \textit{engine:e-sw1} for the \textit{user: shahzad}. It contains \textit{compute, source, destination} and \textit{rate} information that will be used by the engine to process the incoming data from \textit{source}. \eunsungnote{put the example of the configuration file as a figure, and add explanation.}\shahzadnote{done}
The DPI engine, written in C language, is the core of EPB. \ignore{It is written is C language.} It takes raw packets as input and outputs the processed packets based on the settings provided in the configuration file. 
This DPI engine in involved in supporting user requirements, i.e., data type, rate, jitter and arithmetic computations. 
\ignore{Previously attempts have been made by researchers to combine SDN with EPBs.} \ignore{Udechukwu \textit{et al}.\eunsungnote{Use first author's last name} \cite{Extendingopenflow} uses a DPI engine as EPB in addition to an extended OpenFlow protocol to apply traffic engineering techniques in SDN.} \eunsungnote{The cited paper is included in the related work? If so, better remove the above sentence because it looks like being unrelated to our EPB module itself. If not, move to the related work section.}\shahzadnote{it is present in related work. removed the sentence}.

\eunsungnote{it doesn't make sense that SDN module contains base stations in Fig. 4.}\shahzadnote{You are right, changed the figure, omitted the base stations}
\begin{figure}[b]
\centerline{\includegraphics[scale=0.28, trim= 0 0 0 0,clip]{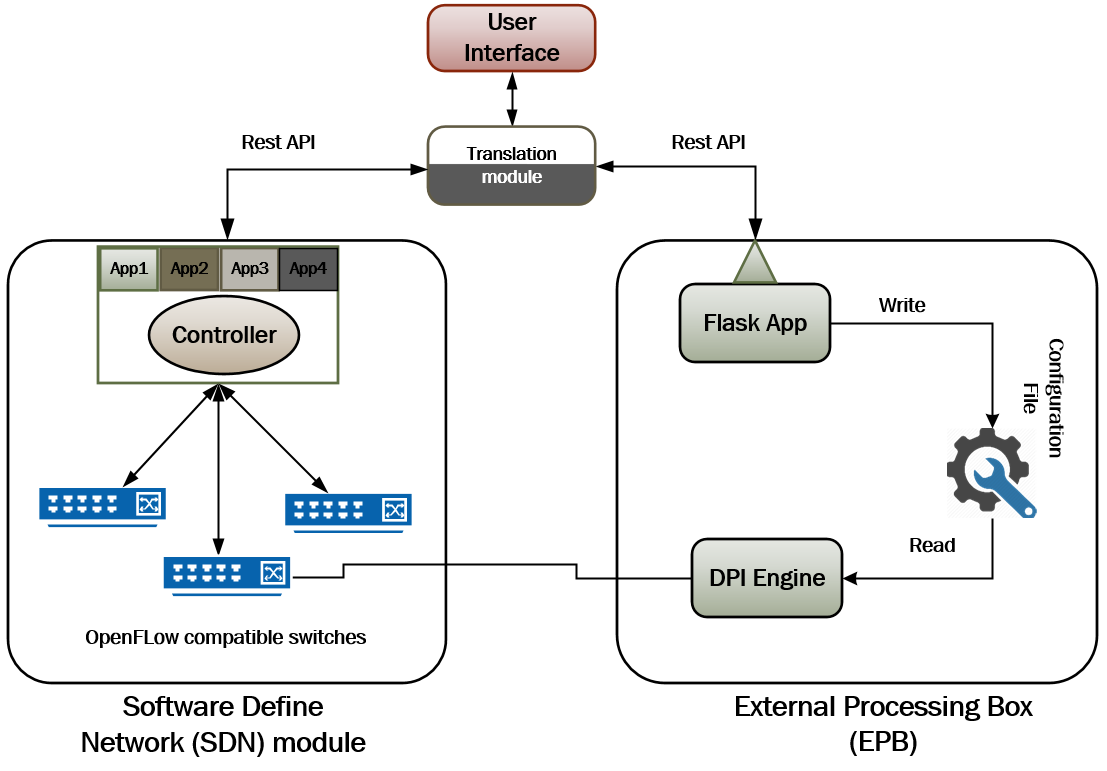}}
\caption{Overall Architecture}
\label{fig-architecture}
\end{figure}
\subsection{How it works}
A user sends requests to FLIP using a script like syntax (see Section \ref{syntax})\eunsungnote{take care of ??}. The request contains the user requirement parameters, and the request is passed to the Translation module. 
\eunsungnote{The following paragraph has redundant sentences. Rewrite like "The translation module convert... Next, the module..."}\shahzadnote{Rewritten}
The Translation module converts the user request into appropriate REST requests. These REST requests are then sent to the designated module (SDN, EPB). Next, the Translation module generates the network graph and Steiner tree for datapath selection and installation, which is discussed in Section \ref{datapath installation}.

For example a user wants to collect data from base stations \textit{bs1} and \textit{bs2} in Fig. \ref{fig-target-network} and wants to perform $sum$ operation on the incoming data, he can write a command as below. \eunsungnote{Needs a brief description of the below command because user command syntax has not been explained yet. For example, when a user want to collect data from ...and do operations, he can write a command as below.}\shahzadnote{Changed the above sentences}
\begin{align*}
    datapath\_m(&bs1,bs2, switch\leftarrow sw,computation\leftarrow sum,\\
    &destination\leftarrow dest)
\end{align*}
This request will establish datapath between $bs1$, $bs2$, the DPI engine connected to $sw$ and $dest$. We have used the Networkx Steiner Tree library in FLIP to generate datapath in the network \cite{networkx}. The request is broken into multiple REST requests. Some of which go to the SDN module and others to the DPI engine. Selection of the DPI engine can be manual as above or automated. We describe the details in Section \ref{syntax}. 

For the given example, the SDN module gets the information about the source nodes (bs1, bs2), destination, and selected DPI engine for datapath installation. Switch $sw$ also receives a REST request to redirect the traffic, which is destined from source nodes to the destination, to the selected engine for computation. The SDN module in association with the translation\eunsungnote{be consistent with module name. Capital letter? with "the"?}\shahzadnote{Capital ``T" because it is start of a sentence. What am I missing?} module serves two major tasks; one is to provide the network detail (i.e., hosts, links, topology, etc.) to the Translation module, and the other is to install a datapath for the given source nodes, switch, DPI engine, and destination. 
There are three applications running on top of the SDN controller for this. 
\begin{itemize}
\item OpenFlow control application (ofctl\_rest): provides REST APIs for retrieving and updating the switch statistics. 
\item The topology discovery application (rest\_topology): provides REST APIs for discovering hosts, links, and switches in the network.  \item Controller basic application (routing\_ryu): sends traffic to the controller when there is no table entry for a given flow of traffic. It also complements the rest\_topology in the initial discovery of hosts. \eunsungnote{The role is redundant with topology discovery application. If this is trivial, comment out the sentence for clarity, or rewrite the sentence.}\shahzadnote{the topology app does not return the hosts until we have such app also running. it has basic functionality but it is not trivial. Sentence rewritten.}
\end{itemize}

For the given example, EPB gets the information about source nodes, destination, and computation (i.e., sum) . This information is passed by the Translation module to the flask web application in the form of REST requests. Next, the flask web application writes this information into the engine configuration file that contains all the configuration for the selected DPI engine. \eunsungnote{this cause some confusion. In the above sentence, you write "gets the source nodes, destination, and computation (i.e., sum) information." Merge two sentences and write a clear one sentence.}
\eunsungnote{The following sentence is also unclear. engine receives the info, then flask access config file?}\shahzadnote{rewritten again}Next, the selected DPI engine reads this configuration file to retrieve its configuration and processes packets accordingly. 
In this example, DPI engine\eunsungnote{What is it? unclear.}\shahzadnote{it$\rightarrow$DPI engine} has to compute $sum$ on data from source nodes (bs1,bs2). 

\eunsungnote{rewrite the following. target network subsection is relocated.}\shahzadnote{rewritten}
\section{User Command Syntax for Script} \label{syntax}
FLIP aims at allowing a user to set up the IoT environment in an automated and distributed manner, but there may be a time when the user wants to interact with a particular module, or only some of the configuration needs to be changed. To address such scenarios, we divided the syntax into two categories to make our framework flexible.
\begin{itemize}
    \item Syntax for automated network setup
    \item Syntax for manual network setup
\end{itemize}

The data structure used in FLIP is JSON. So all the requests return data in JSON format which can be utilized in a serialized manner by combining multiple requests together. 

\subsection{Syntax for automated network setup}
With the syntax for automated network setup, a user only has to specify source nodes\eunsungnote{you need to be precise. base stations? or IoT devices?}\shahzadnote{base station$\rightarrow$source nodes. In reference to the assumption in coverage section} and requirements and does not have to worry about the engine location or datapath selection. For this, FLIP has the following syntax: \eunsungnote{The following syntax has several problems. 1) \{\} not \(\)? 2) looks like for a group of sources only one requirement. 3) only three operations supported? 4) you have to mention that nested operations are possible.}\shahzadnote{changed as suggested}
\begin{align*}
    &datapath\_a(operation(sources\  | \ list*), destination\leftarrow d,\\
    &\qquad \qquad \quad \  requirement\leftarrow r*) \\
    &operation\leftarrow [min,\ max,\ sum,\ sub,\ mul]\\
    &sources\leftarrow \{a\ set\ of\ source\ nodes\}\\
    &list\leftarrow [(,operation(sources\ |\ list*),)]
\end{align*}

\subsection{Syntax for manual network setup}
Table \ref{tab1} shows the commands for the manual network setup. The commands are categorized into three classes: topology, datapath, and DPI engine. The syntax for manual network setup is as follows. \eunsungnote{Think about the syntax definition again.}\shahzadnote{gave it a try}
\begin{align*}
    &datapath\_m(operation(sources), destination\leftarrow d,\\
    &\qquad \qquad \quad \  requirement\leftarrow r*) \\
    &operation\leftarrow [min,\ max,\ sum,\ sub,\ mul,\ list]\\
    &sources\leftarrow \{a\ set\ of\ source\ nodes\ |\ switch\ |\\
    &\qquad \qquad \quad \ DPI\ engine\}
\end{align*}

\begin{table}[htbp]
\caption{Syntax description}
\begin{center}
\begin{tabular}{| c | m{4cm}| c | }
\hline
\textbf{Commands}&\textbf{Description}&\textbf{Category}\\
\hline
\textit{getswitches} & Get switches information- dpid, port, mac,ip etc. & Topology\\
\hline
\textit{getlinks} & Get links information in the network. & Topology\\
\hline
\textit{gethosts} & Get connected hosts information- mac, ip, switch port etc. & Topology\\
\hline
\hline
\textit{getswdesc} & Get switch description for a given dpid e.g. hw\_desc, sw\_desc, manufacturer etc. & Datapath\\
\hline
\textit{getflows} & Get packes flows information for a given dpid. & Datapath\\
\hline
\textit{gettables} & Get all tables for a given dpid. & Datapath\\
\hline
\textit{getports} & Get all ports information for a given dpid. e.g. received packet, transfered packets, dropped(rx, tx) etc. & Datapath\\
\hline
\textit{addflow} & add a new flow entry for a given dpid. & Datapath\\
\hline
\textit{modflow} & Modify existing flows for a given dpid. & Datapath\\
\hline
\textit{delflow} & delete a specific flow for a given dpid. & Datapath\\
\hline
\textit{delflowall} & delete all flows for a given dpid. & Datapath\\
\hline
\hline
\textit{getconfig} & Get configuration details for a given DPI engine. & DPI engine \\
\hline
\textit{getconfig/user} & Get configuration details of \textit{user} for a given DPI engine.& DPI engine\\
\hline
\textit{setconfig/user} & Set configuration details of \textit{user} for a given DPI engine.& DPI engine\\
\hline
\textit{setconfig/user/module} & Set configuration details of \textit{module} of \textit{user} for a given DPI engine.& DPI engine\\
\hline
\hline
\textit{datapath\_m} & Get topology info, set datapath and update configuration for given hosts and DPI engine. & All\\
\hline
\end{tabular}
\label{tab1}
\end{center}
\end{table}

\subsection{Example}\label{syntax-example}
Let's say the user wants to perform the following operations on the selected base stations in the target network in Fig. \ref{fig-target-network} in an aggregated manner. Eq. \ref{eq:1} show the desired request in a general syntax.
\eunsungnote{If we get base stations as input, somewhere in the paper, we have to make assumptions such as we collect all sensor data from a base stations. No way to select a specific IoT device. Or explain as a proof of concept we select base stations in our paper, but can be extended easily to include specific IoT devices.}\shahzadnote{Right, since we introduced translation in coverage, I have added the assumption at the end of coverage section. I think adding proof for specific device access at this time is not feasible.}
\begin{align}\label{eq:1}
    \begin{aligned}
    max(&avg(bs1:bs10),avg(bs11:bs100), \\
    &max(min(bs101:bs200), min(bs201:300)))
    \end{aligned}
\end{align}

\begin{figure}[t]
\centerline{\includegraphics[scale=0.34, trim= 0 0 0 0,clip]{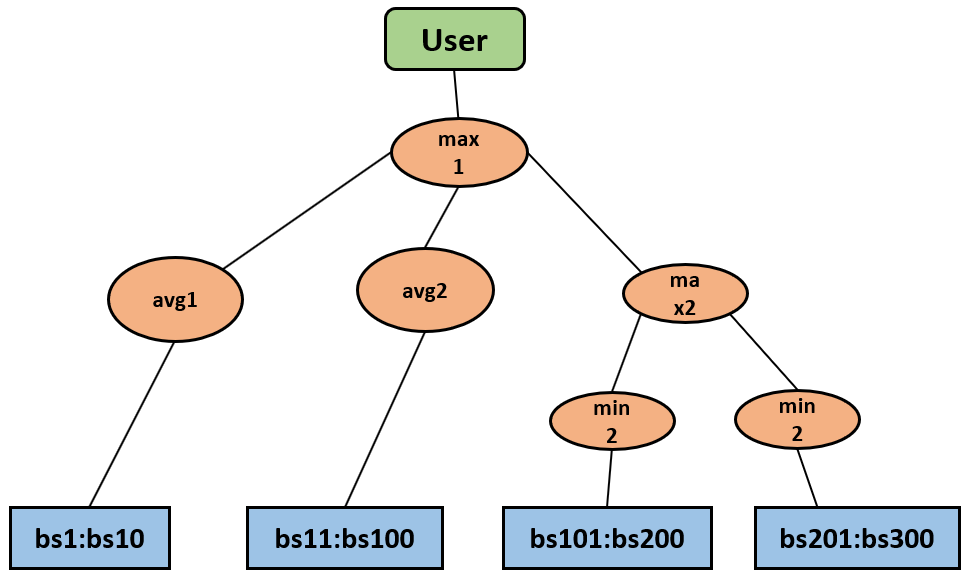}}
\caption{Task graph for Eq. \ref{eq:1}}
\label{fig-taskgraph}
\end{figure}

In Automated syntax case, it can be written like this.
\begin{align*}
    datapath\_a(&max(avg(bs1:bs10), avg(bs11:bs100),\\
    &max(min(bs101:bs200),min(bs201:bs300))),\\ &destination\leftarrow user)
\end{align*}
The Translation module builds a task graph (Fig. \ref{fig-taskgraph}) from this request. This task graph is than mapped to the actual network graph (Fig. \ref{fig-target-network}) to assign DPI engines from the network graph to each operation in the task graph in an automated and distributed manner. Heuristics for this mapping are presented in Section \ref{datapath installation}. Once the heuristics map the operations to DPI engines then a Steiner Tree is created between the Source nodes, DPI engines and destination (user). The task graph has the following components.
\begin{itemize}
    \item S: Set of base stations (bsxx)
    \item OP: Operation nodes (max1,avg1,avg2,max2,min1,min2)
    \item D: Destination node (user)
\end{itemize}

For the manual syntax case this request can be broken down into a multi-part request as follows: 
\begin{align*}
    \begin{aligned}
    &datapath\_m(\{bs201:bs300\}, switch\leftarrow sw4,\\
    &\qquad \qquad \ \quad compute\leftarrow min, destination\leftarrow sw5[engine])\\
    &datapath\_m(\{bs101:bs200\}, switch\leftarrow sw3,\\
    &\qquad \qquad \ \quad compute\leftarrow min, destination\leftarrow sw5[engine])\\
    &datapath\_m(sw4[engine], sw3[engine], switch\leftarrow sw5,\\
    &\qquad \qquad \ \quad compute\leftarrow max, destination\leftarrow sw3[engine])\\
    &datapath\_m(\{bs11:bs100\}, switch\leftarrow sw2,\\
    &\qquad \qquad \ \quad compute\leftarrow avg, destination\leftarrow sw3[engine])\\
    &datapath\_m(\{bs1:bs11\}, switch\leftarrow sw1, compute\leftarrow avg,\\
    &\qquad \qquad \qquad destination\leftarrow sw3[engine])\\
    &datapath\_m(sw1[engine],sw2[engine],sw5[engine],\\
    &\qquad \qquad \qquad switch\leftarrow sw3, compute\leftarrow max,\\
    &\qquad \qquad \qquad destination\leftarrow user])
    \end{aligned}
\end{align*}
As we saw in previous $datapath\_m$ example, For each $datapath\_m$ command here, the Steiner tree will be created between source nodes, switch, selected DPI engine and the destination. The difference between automated and manual can be clearly seen. In the automated case, the user did not provide any switch or DPI engine or mapping of operations to switches. while in the manual case, the user provided this detail for each $datapath\_m$ command.

\section{Flexible IoT datapath installation using REST APIs} \label{datapath installation}
In SDN there are two approaches for setting up datapath in the network, proactive approach and reactive approach. While the proactive approach is easy to establish, the reactive approach is more complex and involves multiple dependencies. In this paper we implement proactive approach, referred to as \textit{static datapath installation}, using an SDN application. For the reactive approach, referred to as \textit{dynamic datapath installation}, a flexible datapath installation for the IoT is shown using \textit{Steiner tree}. 
\subsection{Static datapath installation}
In this approach all the forwarding devices (switches) are programmed in advance i.e. table entries are populated beforehand to forward traffic between all the connected hosts. This is an efficient method because the traffic to the controller is minimized, saving a lot of overhead. But it becomes inefficient when there are changes made into the network e.g. a link gets down, a host is unreachable etc.

This is referred to as the Static datapath installation in this paper and is achieved by having a controller application populate all the switches as the network is instantiated.
\subsection{Dynamic datapath installation}
This approach is about reacting to changes in the network and maintain connectivity. It may also be the case that the user wants to establish a datapath for a specific hosts or DPI engines as we saw examples in previous sections. So whenever there is a change detected in the network, the controller in notified and a solution is generated on the go. This approach is better suited in conditions where the user wants the network to be able to change or adapt to changes as is common in most IoT use cases. 

This is referred to as the dynamic datapath installation in this paper. Our Translation module in association with the SDN module is responsible for the dynamic installation of table entries in the switches according to the user requirements. Whenever a $datapath$ request is made by the user. The Translation module sends an initial request for the topology discovery that is executed by the topology application running on the SDN controller. In reply, the module gets the information about the hosts, links and switches in the network. Using this data combined with the user provided information (Source nodes, destination etc.), the Translation module generates a network graph using python Networkx library. Next, Steiner Tree is created between source nodes, switches, DPI engines and destination.

In case of manual $datapath$, The Steiner Tree is created with the information provided by the User. While in case of Automated $datapath$, mapping is performed between task graph (Fig. \ref{fig-taskgraph}) and the actual network graph (Fig. \ref{fig-target-network}). Following subsection explains the heuristics for the dynamic datapath selection and mapping using Steiner Tree. 

Once the path computation is complete. The Translation module starts sending REST requests for the dynamic datapath installation. These requests are received by the ofctl\_rest application running on the SDN controller. The user is notified for the successful installation of all the rules in the respective switches. Then the  network topology can be displayed using python matlab library.
\begin{algorithm}[t]
\SetAlgoLined
\caption{Heuristic algorithm to find Steiner Tree }
\label{algoritm}
\SetKwInOut{Input}{Input}
\SetKwInOut{Output}{Output}
\textbf{Definitions:}\\
N = (V, E, w) : network topology,\\
V: a set of nodes, three kinds of nodes (base station node, switch node, destination node).\\
E: links between nodes.\\
w: weight of a link, which is assumed to be delay in our paper.\\
$S = \{s \mid s \in V, \text{s is a base station}\}$ A set of base station nodes.\\
D: a destination server/host\\
T: a tree describing user requirements. root is a destination, leaf nodes are base stations (S), and intermediate nodes are operation nodes.\\
M: a set of switch nodes on which operation nodes are assigned.\\
\Input{N, S, D, and T}
\Output{ST: Steiner tree for the efficient datapath}
\Begin{
    $edgenodes[] \leftarrow leafonlyparents(T)$\;
     \tcc{returns nodes with only leaf children}
     \If{$edgenodes \neq NULL$}{
    \For{ $node \in  edgenodes$}
    {
        $leafnode \leftarrow leftchild(node)$\; 
        \tcc{returns left most child for node}
        $edgeswitch[node] \leftarrow connectedsw(leafnode,N)$\; 
        \tcc{returns connected switch}
        $pnode \leftarrow parent(node)$\;
         \While{$pnode \neq NULL$}
        {
          \If{$pnode \in visited$}
              {break loop\;}
             $visited \leftarrow add.pnode$\;
             $interswitch[pnode] \leftarrow adjswitch(edgeswitch[node],N)$\;
             $pnode \leftarrow parent(pnode)$\;
        }
    }
    }
    $M \leftarrow \{edgeswitch[] + interswitch[]\}$\;
    $ST \leftarrow steinertree(S,M,,D,N)$\;
}
\end{algorithm}

\subsubsection{Heuristics to find Steiner tree}
In a given network topology, the user may want to establish datapath between a certain number of hosts as shown in examples in the preceding sections. Since the IoT network topologies contain a large number of nodes in general, it is better suited to create a Steiner tree among the desired nodes (source base stations, switches, DPI engines and destination) and establish datapath between them to reduce network resource utilization.
The user provides source base stations ($S$), computation information, and destination $D$. We use a heuristic algorithm to find optimal DPI engines for computation and then create a Steiner tree based on Networkx library. This algorithm returns an approximation to the Steiner tree for $S$ and $D$ with optimally selected switches linked to DPI engines.  Algorithm~\ref{algoritm} details the applied heuristics.

The input to the Algorithm is $N$- the network topology, $S$- source base stations, destination $D$ and $T$- a Tree that shows the task graph representation of user requirements with root $D$, leaf nodes $S$ and intermediate nodes as operation nodes. The Output is an approximation of Steiner tree whose weight is within a $(2 - (2 / T))$ factor of the weight of the optimal Steiner tree \cite{networkx}.
First the edgenodes are collected. These are the operation nodes that are directly connceted to $S$. Then for each edgenode, a single leafnode is iterated (i.e. one leaf node for each parent edgenode). Also keep in mind that each leafnode $\in S$. With this leafnode we create a dictionary $edgeswitch$ that is populated by calling the $connectedsw(node,topology)$ function for each leafnode. The $connectedsw()$ funtion returns the switch id ($dpid$) that is connected to $leafnode$. Eventually the $edgeswitch$ dictionary contains the relations between edgenodes (first level operation nodes) of $T$ and the edge switches (directly connected to $S$) of $N$. Then we move to $pnode$-parent of iterating $node$ in \textbf{for} loop. A $while$ loop keeps iterating until $pnode \neq NULL$. $pnode$ is checked against a $visited$ list that keeps track of the visited parent nodes. If $pnode$ is not visited, add $pnode$ to the $visited$ list otherwise break the $while$ loop. Then we create another dictionary called $interswitch$ that maps each $pnode$ to a $switch$ in $N$ by calling the $adjswitch(dictionary[key],topology)$ function. This function returns the adjacent switch in topology $N$ for the provided key value i.e. values from the previous dictionary $edgeswitch$. Then we increment $pnode$ i.e. move to it's parent node and repeat the process. Once the $for$ loop is complete, a list $M$ combines $edgeswitch$ and $interswitch$ dictionaries. This $M$ list now contains the mapped switches in $N$ to each operation node in $T$. Now we call the $steinertree()$ with inputs $S$, $M$, $D$, and $N$.   

\section{Experimental Evaluation} \label{Experiments}
\begin{figure*}[htbp]
\centerline{\includegraphics[scale=0.37, trim= 0 0 0 0,clip]{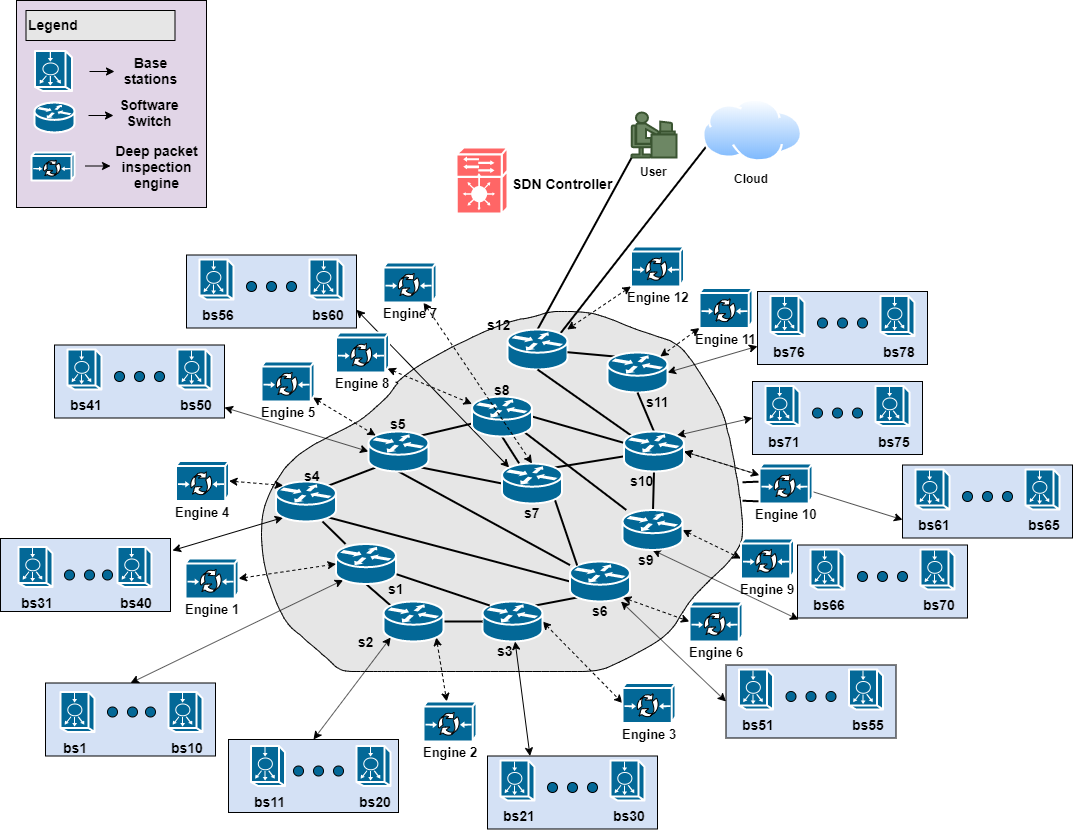}}
\caption{Mininet Experimental Network}
\label{fig-experiment}
\end{figure*}

As a proof of concept we tested our framework on mininet which is an emulation environment to build and test network topologies. Mininet provides the option to add software switches and remote controller to support SDN related experiments. We created a partial mesh topology with:
\begin{itemize}
    \item 1 RYU SDN controller
    \item 12 OVS switches
    \item 78 Base stations
    \item 12 DPI engines
    \item 1 Destination (User),
    \item 1 Cloud node
\end{itemize} 
The topology is shown in Fig. \ref{fig-experiment}. The SDN controller connections are omitted in the diagram to maintain clarity. 

\begin{figure}[htbp]
\centerline{\includegraphics[scale=0.48, trim= 0 0 0 0,clip]{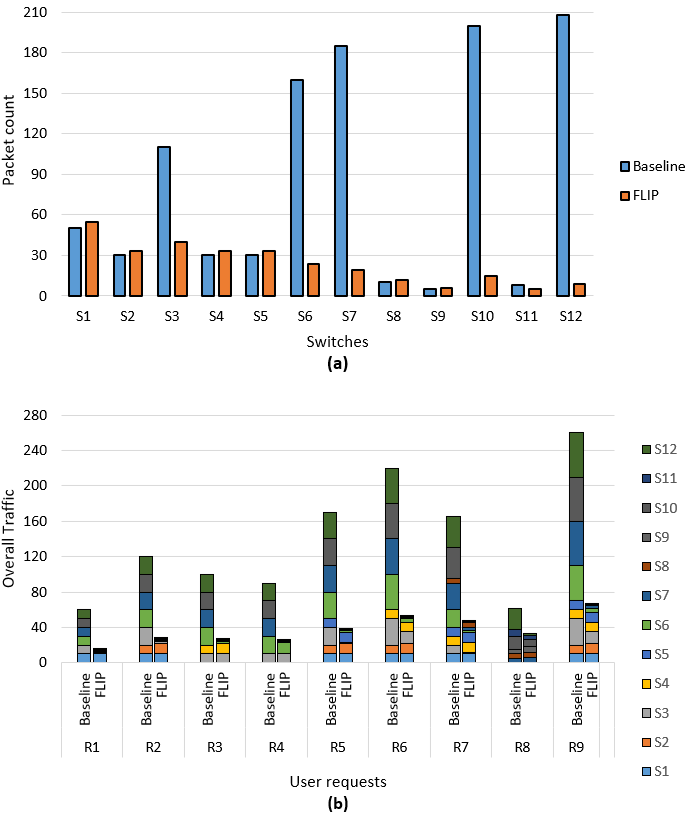}}
\caption{Experiment results: a) Packet count for each switch b) Overall network traffic for each request}
\label{fig-results}
\end{figure}

For baseline comparison we have used the same topology without the DPI engines and FLIP intervention i.e. all packets are sent to the user. We send 9 different user requests to this experimental network with and without FLIP's intervention. We have assumed that without FLIP the request is computed at the destination node. The requests are enumerated below.
\begin{align*}
    &R1\rightarrow max(bs1:bs10)\\
    &R2\rightarrow avg(max(bs1:h10),max(bs11:bs20))\\
    &R3\rightarrow avg(min(bs21:bs30),min(bs31:bs40))\\
    &R4\rightarrow sum(avg(bs21:bs30),avg(bs51:bs60))\\
    &R5\rightarrow sum(max(bs1:bs10),max(bs11:bs20),\\
    &\qquad \qquad \quad max(bs41:bs50))\\
    &R6\rightarrow sum(max(bs1:bs10),max(bs11:bs20),\\
    &\qquad \qquad \quad min(bs21:bs30),min(bs31:bs40))\\
    &R7\rightarrow max(max(bs1:bs10),min(bs31:bs40),\\
    &\qquad \qquad \qquad max(bs41:bs50),min(bs56:bs60))\\
    &R8\rightarrow max(avg(bs56:bs60),avg(bs61:65),\\
    &\qquad \qquad \quad max(min(bs66:bs70),min(bs71:bs75)),\\
    &\qquad \qquad \quad max(bs76:bs78))\\
    &R9\rightarrow max(avg(bs1:bs10),avg(bs11:bs20),\\
    &\qquad \qquad \quad max(min(bs21:bs30),min(bs31:bs40)),\\
    &\qquad \qquad \quad max(bs41:bs50))\\
\end{align*}
We have generated two types of results based on the above requests. In Fig. \ref{fig-results}(a) we have compared the packet count on each switch for all the requests collectively. The packets count is obtained by sending REST request to each switch with a flow filter for packets destined to user (destination node). It can be observed that apart from the edge switches (switches directly connected to concerned base stations) the network utilization is reduced because with FLIP enabled, only processed results travel to the user instead of all the packets. In Fig. \ref{fig-results}(b) overall network traffic for each request is shown. For each request we compare FLIP with baseline for the overall network utilization. From these results it can be observed that FLIP can reduce the overall network utilization by $45\% \sim 77\%$. This shows that FLIP has the potential to be effectively used in large-scale IoT environments.

\section{Related Work} \label{Related-work}
In the past decade, many techniques and frameworks have been proposed to make programming, routing and configuration of IoT systems easy and more efficient. Aggregate programming \cite{aggregate} uses field calculus to program the IoT fabric as a whole such that the basic unit is no longer a single device. WoTT \cite{wott} is an application oriented IoT test-bed that focuses on interoperability issues in IoT but it does not support streaming data. PHEME \cite{pheme} is an analytics re-purposing system that re-purposes Web analytics for IoT data collection. The distinction between our framework and the above is that our framework is SDN based, supports streaming data and has DPI capabilities.

Some other related research works include Fluidware \cite{fluidware}, Fibbing \cite{fibbing}, Muppet \cite{muppet}. Fluidware is similar to \cite{aggregate} in terms of programming the IoT devices based on the service and application rather than the device itself but no experimental implementation has been provided. Also it is not SDN based. Fibbing is similar to our framework in some aspects. It has similar script language style, it uses SDN as a base architecture. The difference is that it focuses on combining SDN with Tradition network protocols like IGP and BGP instead of IoT protocols. Muppet is an edge-based multi-protocol architecture for large-scale IoT deployment and service automation. It focuses on the interoperability issue in the IoT and proposes an SDN based (P4) solution to connect heterogeneous devices. While our framework focuses more on the datapath generation for large-scale streaming data and packet processing. 
Other distinct features of FLIP include:
\begin{itemize}
    \item Multi user support
    \item Multi stream support for a given user
    \item Central Rest Communication 
    \item Multiple engines deployment
    \item Multiple SDN applications for communication and routing 
\end{itemize}

\section{Conclusion}\label{conclusion}
In this paper, we present FLIP - FLexible IoT Path Programming Framework for large-scale IoT. Moving away from the centralized cloud based IoT automation and processing, FLIP introduces efficient in-network processing of big IoT data that reduces the amount of data arriving at server/cloud. We show through experiments that FLIP has the potential to reduce the bottleneck issues of handling massive IoT data at the server/cloud while meeting various user requirements. Experimental results show that the overall network utilization is reduced by $45\%\sim 77\%$. 
\ignore{
\section*{Acknowledgment}
The preferred spelling of the word ``acknowledgment'' in America is without 
an ``e'' after the ``g''. Avoid the stilted expression ``one of us (R. B. 
G.) thanks $\ldots$''. Instead, try ``R. B. G. thanks$\ldots$''. Put sponsor 
acknowledgments in the unnumbered footnote on the first page.}

\bibliographystyle{unsrt}
\bibliography{refs}

\end{document}